\documentclass[conference]{IEEEtran}
\IEEEoverridecommandlockouts

\usepackage{cite}
\usepackage{amsmath,amssymb,amsfonts}
\usepackage{algorithmic}
\usepackage{graphicx}
\usepackage{textcomp}
\usepackage{xcolor}
\usepackage{booktabs}
\usepackage{multirow}
\usepackage{orcidlink}
\usepackage{tikz}
\usepackage{comment}
\usetikzlibrary{arrows.meta, positioning, fit, calc}

\def\BibTeX{{\rm B\kern-.05em{\sc i\kern-.025em b}\kern-.08em
    T\kern-.1667em\lower.7ex\hbox{E}\kern-.125emX}}

\newcommand{\defvspace}{-0.48cm}

\newcommand{\x}{\mathbf{x}}
\newcommand{\m}{\mathbf{m}}
\newcommand{\y}{\mathbf{y}}
\newcommand{\xwm}{\mathbf{\tilde{x}}}

\newcommand{\xwmspoof}{\mathbf{\tilde{x}_{\mathrm{sp}}}}
\newcommand{\devdtw}{s_{\mathrm{DTW}}} 
\newcommand{\seconds}[1]{#1\,\mathrm{s}}

\newcommand{\ms}[1]{#1\,\mathrm{ms}}
\newcommand{\khz}[1]{#1\,\mathrm{kHz}}

\newcommand{\kbps}[1]{#1\,\mathrm{kbps}}
\newcommand{\R}{\mathbb{R}} 
\newcommand{\DTW}{\mathrm{DTW}}
\newcommand{\warpingpath}{\pi^*}
\newcommand{\cosdist}{c_t}

\usepackage[dvipsnames]{xcolor}

\def\BibTeX{{\rm B\kern-.05em{\sc i\kern-.025em b}\kern-.08em
    T\kern-.1667em\lower.7ex\hbox{E}\kern-.125emX}}
\begin{document}
\title{Combining Self-Embedding Audio Watermarking with Ultra-Low-Bitrate Neural Codecs}

\author{\IEEEauthorblockN{Yigitcan {\"O}zer, Xin Wang, Zhe Zhang, Junichi Yamagishi}
\IEEEauthorblockA{National Institute of Informatics, Tokyo, Japan\\
Email: \{yiitozer, wangxin, zhe, jyamagis\}@nii.ac.jp}\thanks{This work was supported by JSPS MEXT KAKENHI Grant 24H00732.}
}

\maketitle

\begin{abstract}
Partial manipulation of speech recordings, where only localized segments of an utterance are altered, poses a significant challenge for content integrity verification, as reliable detection and localization of such edits becomes harder as the manipulated proportion decreases.
Watermarking offers a proactive defense alternative by embedding auxiliary information prior to distribution; classical hash-based schemes achieve near-perfect detection and localization under ideal conditions, but the original content cannot be recovered once a segment is manipulated.
Building on a prior self-embedding audio steganography framework, this work presents an initial exploration of proactive defense performance under ideal conditions, extending the investigation along three axes: frame-level localization, multi-bit least significant bit variants, and evaluation across multiple ultra-low-bitrate neural codec representations.
By embedding a compact neural codec representation rather than a cryptographic hash, the framework additionally enables recovery of the manipulated regions, while supporting training-free detection and localization without spoofed examples.
Experiments across four controlled manipulation types under ideal channel conditions show that the embedded payload, and hence an approximate reconstruction of the authentic content, is always fully recovered without bit errors. 
The results also indicate that the choice of neural codec is the dominant factor for detection and localization performance.
\end{abstract}

\begin{IEEEkeywords}
audio watermarking, partial deepfake detection, self-embedding watermarking, self-recovery watermarking
\end{IEEEkeywords}

\section{Introduction}
\label{sec:intro}
Partial manipulation of speech recordings poses a fundamental challenge for audio integrity\,\cite{HeEtAl25_ManipulatedRegionsSurvey_arXiv}. 
Unlike fully synthesized speech, such manipulations preserve most of the original recording while altering only a localized segment\,\cite{ZhangEtAl23_PartialSpoof_TASLP}.
Proactive defense strategies, e.g.,\, watermarking, address this by embedding auxiliary information into the signal prior to distribution\,\cite{BenderEtAl96_DataHiding_IBM}, enabling subsequent integrity verification.

In classical watermarking, fragile, semi-fragile, and robust watermarks serve different security goals\,\cite{SteinebachDittmann03_AudioAuthentication_EURASIP}.
Robust watermarks are engineered to survive substantial signal processing and are appropriate for ownership tracing or provenance verification\,\cite{SanRomanEtAl24_AudioSeal_ICML, LiuEtAl24_TimbreWatermarking_NDSS}.
Conversely, precise localization of altered segments rather requires fragile or semi-fragile paradigms. 
Since fragile watermarks are designed to break under the slightest waveform alterations, any local tampering selectively corrupts the embedded payload only within the modified interval.
This sharp contrast between intact and broken payload segments provides a precise spatial indicator of the tampered boundaries under ideal channel conditions. 
Traditionally, these paradigms embed a cryptographic hash of each audio segment, which localizes tampering exactly but cannot recover the original spoken content once a segment has been overwritten\,\cite{HuaEtAl16_TwentyYearsAudioWatermarking_SP, RenzaBL18_AudioAuthenticity_FragileWatermarking_ESA}.
Recently, a neural semi-fragile watermarking approach has been proposed for proactive deepfake speech detection\,\cite{YoonToda25_NeuralSemiFragile_APSIPA}.
However, such learned schemes require training on spoofed examples and cannot recover the original content, which would require embedding the audio itself as the payload.

Self-embedding watermarking, also referred to as self-recovery or self-healing watermarking, embeds an authentication and recovery description derived from the host signal into the carrier signal itself.
This idea has been extensively studied for image and video authentication and recovery\,\cite{FridrichGoljan99_SelfCorrectingImages_ICIP, YogarajahEtAl11_VideoAuthentication_IMVIP, RakhmawatiEtAl20_BlindSelfEmbedding_IJIES, SinghEtAl23_SelfEmbeddingFragileWatermarking_MTA}.
In the audio domain, early speech self-embedding schemes modeled the problem as source--channel coding: a compressed representation of the speech signal, protected by channel coding and accompanied by frame-level hash information, is embedded into the signal, allowing tampered frames to be detected and subsequently reconstructed\,\cite{SarreshtedariEtAl15_SpeechSelfRecovery_TASLP}.
Subsequent work improved embedding transparency via auditory masking\,\cite{MenendezOrtizEtAl18_AudioSelfRecovery_PlosOne} and a further study addressed content replacement with duration changes\,\cite{GomezRicardezGarciaHernandez21_AudioSelfRecovery_Computers}.
These studies establish that audio self-embedding is possible, but they rely on digital signal processing (DSP)-based codecs to generate the recovery payload, e.g.,\,{G.723.1} at $\kbps{6.4}$ and {MELP} at $\kbps{2.4}$ in\,\cite{SarreshtedariEtAl15_SpeechSelfRecovery_TASLP}, and {OPUS} at $\kbps{64}$ in\,\cite{GomezRicardezGarciaHernandez21_AudioSelfRecovery_Computers}.
Since the payload must fit within the strictly limited embedding capacity, the codec bitrate directly governs the achievable redundancy, and hence the extent of manipulation from which the content can be recovered.
Even the lowest-rate standardized {DSP}-based coders, such as {MPEG-4} {HVXC}, operate at $\kbps{2.0}$\,\cite{NishiguchiEtAl98_HVXC_ICCE}, whereas recent neural audio codecs have pushed speech coding below $\kbps{1}$ while preserving intelligible, speaker-consistent reconstruction\,\cite{MousaviEtAl25_DiscreteAudioTokens_TMLR}, reducing the recovery payload to a fraction of the bitrates used in prior audio self-embedding schemes.

\begin{figure*}[t!]
  \centering
  \includegraphics[width=\linewidth]{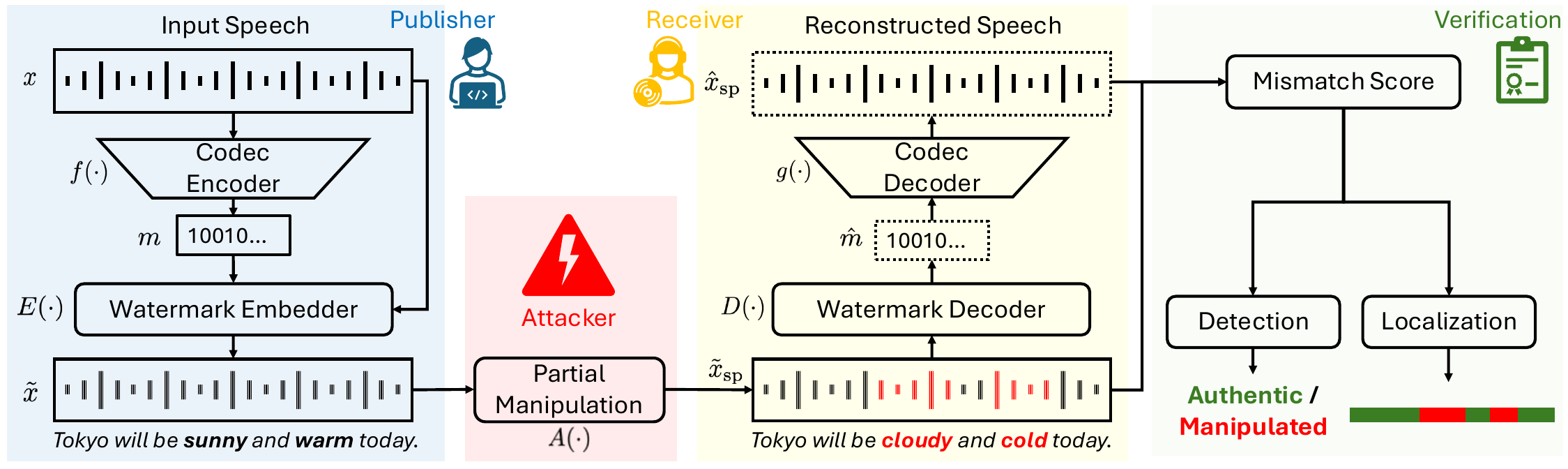}
  \vspace{-0.24cm}
  \caption{Overview of the proposed self-embedding audio watermarking framework. At the publisher side, the carrier signal is encoded by a neural codec and embedded into itself via a watermark embedder. A partial manipulation is applied to the watermarked signal. At the receiver side, the embedded bitstream is recovered by a watermark decoder and passed to a codec decoder to reconstruct the authentic waveform. Finally, a mismatch score between the received signal and its reconstruction is used jointly for utterance-level detection and frame-level localization of manipulated segments.}
  \vspace{\defvspace}
  \label{fig:teaser}
\end{figure*}

In this work, we revisit audio self-embedding in light of recent ultra-low-bitrate neural audio codecs\,\cite{SiuzdakEtAl24_SNAC_NeurIPSWorkshop, LiuEtAl24_SemantiCodec_SP, ParkerEtAl25_ScalingTransformersSpeechCoding_ICLR}.
We embed a compact neural codec representation of the carrier signal into itself; at verification time, the extracted payload is decoded to obtain a self-reconstruction of the authentic speech, and the mismatch between this reconstruction and the received signal is used for detection and localization (see Fig.~\ref{fig:teaser}).
This design preserves the central advantage of self-recovery, which is the ability to reconstruct manipulated content, while allowing us to study whether recent sub-kbps neural codecs provide a practical payload for modern partial speech manipulation scenarios.

Our prior work introduced an initial version of this neural codec self-embedding framework for partial speech manipulation detection\,\cite{OezerEtAl26_AudioStegoPartialDeepfake_Interspeech}, using least significant bit (LSB) embedding\,\cite{BenderEtAl96_DataHiding_IBM} with SNAC as the compression backbone\,\cite{SiuzdakEtAl24_SNAC_NeurIPSWorkshop}.
The present paper extends this framework in three directions:
\emph{(i)}~we extend evaluation from utterance-level detection to frame-level localization;
\emph{(ii)}~we generalize single-bit LSB embedding to multi-bit fragile variants ($k \in \{1,2,4\}$);
\emph{(iii)}~we evaluate the framework across multiple ultra-low-bitrate neural codecs and manipulation types, including replacement, deletion, and insertion.
Experiments show that detection and localization performance are governed primarily by neural codec reconstruction fidelity.

\section{Self-Embedding Audio Watermarking using Neural Audio Codecs}
\label{sec:method}
The proposed framework embeds an ultra-low-bitrate neural codec representation of the carrier signal into itself, enabling a manipulated segment to be approximately reconstructed from the recovered payload as long as a sufficient portion of the watermarked signal remains intact.
This section formalizes the embedding and verification pipeline, details the LSB embedding method, and analyzes the embedding capacity across codec configurations.

\subsection{Problem Formulation}
\label{sec:problem_formulation}
This formulation extends \cite{OezerEtAl26_AudioStegoPartialDeepfake_Interspeech} to multiple bit depths and neural audio codecs.
Let $\x \in \R^T$ denote a discrete-time speech signal of length $T$, sampled at $f_s$ Hz.
An embedder \mbox{$E : \R^T \times \{0,1\}^M \rightarrow \R^T$} conceals a binary message $\m \in \{0,1\}^M$ within $\x$, producing a watermarked signal \mbox{$\xwm = E(\x, \m)$} that remains perceptually indistinguishable from the original.
A decoder \mbox{$D : \R^T \rightarrow \{0,1\}^M$} recovers the message as \mbox{$\hat{\m} = D(\xwm)$}, where $M$ denotes the total embedding budget (in bits) for a signal of length $T$.

The central constraint is that the embedding budget $M$ is strictly limited by imperceptibility, satisfying $M \ll B \cdot T$, where $B$ is the bit depth.
As a result, embedding the waveform $\x$ itself, or even a lossless encoding of $\x$, is infeasible.
More importantly, this constraint precludes redundancy through repeated embedding, which is required for robustness against localized manipulations.

We overcome this limitation by introducing a neural audio codec operating at bitrate $\rho$ (bps) as a compression backbone.
Let \mbox{$f : \R^T \rightarrow \{0,1\}^N$} denote the codec encoder, where \mbox{$N = \rho \cdot (T / f_s)$} is the number of bits in the compressed representation.
For modern ultra-low-bitrate codecs ($\rho < \kbps{1.0}$), the following inequalities hold:
\begin{equation}
    \rho \cdot \frac{T}{f_s} \;\ll\; M \;\ll\; B \cdot T,
    \label{eq:capacity}
\end{equation}
where the left term is the codec representation size, the middle term is the embedding budget, and the right term is the raw waveform size in bits.
Notably, the left inequality implies that the compressed representation $f(\x)$ is sufficiently small to be embedded multiple times within the available embedding budget.
This enables a self-embedding design with explicit redundancy. 
The message is defined as $R$ consecutive repetitions of the codec representation:
\begin{equation}
    \m = (m_1, m_2, \ldots, m_M) = \underbrace{f(\x) \;\|\; \cdots \;\|\; f(\x)}_{R},
    \label{eq:redundant_payload}
\end{equation}
where $\|$ denotes concatenation and $R = \lfloor M / N \rfloor$ is the maximum number of non-overlapping repetitions that fit within the embedding budget $M$. 
Equivalently, the $i^\mathrm{th}$ bit of the $r^\mathrm{th}$ repetition is \mbox{$m_i^{(r)} = m_{(r-1)N + i}$} for \mbox{$i = 1, \ldots, N$} and \mbox{$r = 1, \ldots, R$}, making the repetition structure of $\m$ explicit.
This redundancy enables majority-vote decoding at extraction time, making the recovered codec representation robust to partial manipulations of the watermarked signal.

To model partial manipulations, we introduce a spoofing operator $A(\cdot)$ that modifies localized time intervals $\{\mathcal{I}_k\}_{k=1}^K$, with $\mathcal{I}_k = [t_{\mathrm{start}}^{(k)}, t_{\mathrm{end}}^{(k)}]$, by substitution, insertion, or deletion.
Since $\m$ is embedded within $\xwm$, the manipulation $A(\xwm)$ simultaneously alters the signal content and corrupts the embedded payload within the affected intervals, yielding
\begin{equation*}
    \m' = f(\x) \;\|\; \cdots \;\|\; \textcolor{red}{f'(\x)} \;\|\; \cdots \;\|\; f(\x),
    \label{eq:corrupted_payload}
\end{equation*}
where $f'(\x)$ denotes a repetition corrupted by the manipulation, and intact repetitions $f(\x)$ remain unaffected.
The decoder $D(\y)$ recovers the $N$-bit codec representation \mbox{$\hat{\m} \in \{0,1\}^N$} by majority voting across the $R$ repetitions embedded in $\m'$:
\begin{equation}
    \hat{m}_i = \mathbf{1}\!\left[\frac{1}{R}\sum_{r=1}^{R} 
    m_i^{\prime(r)} \ge 0.5\right], \quad i = 1, \ldots, N,
    \label{eq:majority_vote}
\end{equation}
where $m_i^{\prime(r)}$ denotes the $i^\mathrm{th}$ bit of the $r^{\mathrm{th}}$ repetition in $\m'$, yielding a reliable estimate of $f(\x)$ from the intact copies.
The framework makes no assumptions about the internal mechanism of $A(\cdot)$, and thus covers both TTS-based synthesis and waveform-level splicing operations.

At verification time, let $\y$ denote the received signal, where $\y = \xwm$ in the authentic case and $\y = \xwmspoof = A(\xwm)$ denotes the spoofed signal under partial manipulations.
A codec decoder \mbox{$g : \{0,1\}^N \rightarrow \R^T$} reconstructs the authentic waveform from $\hat{\m}$, yielding the self-reconstruction \mbox{$R(\y) = g(D(\y))$}, and the mismatch score under a dissimilarity measure $d$ is \mbox{$s(\y) = d\big(R(\y),\, \y\big)$}.
Detection is formulated as the binary hypothesis test:
\begin{equation}
\delta(\y) =
    \begin{cases}
    0, & s(\y) \le \tau,\\
    1, & s(\y) > \tau,
    \end{cases}
    \label{eq:detection}
\end{equation}
where $\tau$ denotes a decision threshold, $\delta(\y)=0$ indicates an \emph{authentic} signal, and $\delta(\y)=1$ indicates a \emph{manipulated} signal.

Beyond utterance-level detection, the framework naturally extends to frame-level localization.
By projecting local alignment costs (e.g.,\,from DTW) onto the time axis of $\y$, manipulated regions can be identified directly from the resulting per-frame mismatch profile, as detailed in the next section.

\subsection{Detection and Localization Score via Dynamic Time Warping (DTW)}
\label{sec:score_dtw}
For the proposed self-embedding watermarking approach, detection and  localization scores are derived from the DTW alignment between the received signal $\y$  and its self-reconstruction $R(\y) = g(D(\y))$ (see \S\,\ref{sec:problem_formulation}).
DTW provides a mechanism for aligning two time sequences under possible local temporal deviations\,\cite{Mueller07_DTW_IRMM}. 
It is particularly suitable in our setting, as partial speech manipulations introduce temporal inconsistencies between $\y$ and $R(\y)$.
By allowing non-linear temporal alignment, DTW separates structural mismatches from trivial timing offsets.
We apply DTW to normalized log-mel spectrogram features, following\,\cite{OezerEtAl26_AudioStegoPartialDeepfake_Interspeech}.

Let $\warpingpath$ denote the optimal warping path obtained by $\DTW(R(\y), \y)$, and let $\cosdist$ denote the local cosine distance between the two feature sequences at step $t$ along $\warpingpath$.
The utterance-level detection score is defined as the mean of the smoothed per-step costs:
\begin{equation*}
    \devdtw(\y) = \frac{1}{|\warpingpath|} \sum_{t \in \warpingpath} \cosdist,
    \label{eq:devdtw}
\end{equation*}
where higher values indicate a higher likelihood of manipulation.

For frame-level localization, the local cost $c_t$ is projected back onto the time axis of the received signal $\y$ by aggregating costs per target frame along $\pi^*$.
The resulting per-frame cost profile is smoothed and compared against binary ground-truth (GT) labels derived from word-level boundaries for frame-level localization evaluation.

\subsection{Capacity and Repetition Analysis}
\label{sec:capacity}
A key design parameter of the proposed framework is the number of repetitions $R$ of the codec representation that can be embedded within the available capacity, as this directly determines robustness to partial manipulation under majority-vote decoding.
Let $C$ denote the embedding capacity in bps of the chosen method, related to the total embedding budget $M$ in \S\,\ref{sec:problem_formulation} by $M = C \cdot T / f_s$.
Accordingly, we express all payload quantities in bps when analyzing repetition rates.
The proposed framework is agnostic to this choice, and any embedding method satisfying the capacity constraint in \eqref{eq:capacity} can serve as a drop-in replacement.
The embedding method carries the same self-referential payload $\m$ defined in \eqref{eq:redundant_payload}, where each repetition of the codec representation $f(\x)$ is prepended with a $64$-bit synchronization preamble.
The number of repetitions per second is therefore \mbox{$R = \lfloor C / (\rho + 64) \rfloor$}, where $\rho$ accounts for the codec representation size and the additional $64$ bits for the synchronization preamble.
The specific values of $C$ for each embedding method are derived in \S\,\ref{sec:embedding_methods}, and the codec-specific values of $\rho$ are detailed in \S\,\ref{sec:lb_codecs}.

\subsection{Least Significant Bit (LSB) Embedding}
\label{sec:embedding_methods}
In this work, we instantiate the framework with LSB encoding.
%
Rather than applying LSB in its standard form, we adopt a temporally repetitive embedding strategy that enables robust payload recovery even when portions of the watermarked signal are manipulated, as detailed in the following.

LSB embedding encodes information by replacing the $k$ least significant bits of each audio sample with payload bits, where $k \in \{1, 2, 4\}$ controls the capacity--distortion tradeoff.
For a $k$-bit scheme, the $k$ consecutive payload bits  $(m_{ki}, \ldots, m_{ki+k-1}) \in \{0,1\}^k$ are packed into  the $k$ least significant bits of the audio sample $x_i$, introducing a maximum perturbation of $2^k - 1$ quantization steps per sample.
The embedding capacity scales linearly with $k$ as $C = k \cdot f_s$ bps; i.e.,\,at $f_s = \khz{16}$, the three variants yield capacities of $16{,}000$, $32{,}000$, and $64{,}000$ bps for $k = 1, 2, 4$, respectively\footnote{The $k=1$ case recovers the standard single-bit LSB scheme evaluated in our prior work\,\cite{OezerEtAl26_AudioStegoPartialDeepfake_Interspeech}.}.

To account for partial manipulations to the watermarked signal, the codec representation $f(\x)$ is embedded repeatedly across the full duration of the carrier, as explained in \S\,\ref{sec:problem_formulation}.
The $r^\mathrm{th}$ repetition begins at sample index \mbox{$r \lceil N / k \rceil$}, yielding $R$ non-overlapping copies of the payload.
At extraction time, each bit is recovered via majority voting across repetitions as defined in \eqref{eq:majority_vote}, yielding a reliable estimate of $f(\x)$ from the intact copies even when some repetitions are corrupted by partial manipulations.

\subsection{Ultra-Low-Bitrate Neural Audio Codecs}
\label{sec:lb_codecs}
We evaluate three openly available ultra-low-bitrate neural codecs in this work.
SNAC\,\cite{SiuzdakEtAl24_SNAC_NeurIPSWorkshop} extends the RVQ framework by introducing quantizers operating at multiple temporal resolutions, achieving $\kbps{0.98}$ for speech at $\khz{24}$.
SemantiCodec\,\cite{LiuEtAl24_SemantiCodec_SP} adopts a fundamentally different architecture, decoupling semantic and acoustic information across two VQ layers and employing a latent diffusion model as the decoder\,\cite{LiuEtAl23_AudioLDM_ICML}, reaching bitrates as low as $\kbps{0.31}$.
The Transformer Audio AutoEncoder (TAAE)\,\cite{ParkerEtAl25_ScalingTransformersSpeechCoding_ICLR} departs from the convolutional paradigm by scaling a transformer-based encoder--decoder to approximately $950$M parameters and  replacing RVQ with Finite Scalar Quantization (FSQ)\,\cite{MentzerEtAl24_FSQ_ICLR}, achieving $\kbps{0.4}$ and $\kbps{0.7}$ for speech at $\khz{16}$.
However, FSQ indices can exceed the \texttt{uint16} range.
Consequently, each token requires \texttt{int32} storage ($32$ bits).
The actual embedded payload sizes therefore differ substantially from the nominal bitrates, as reflected in Table~\ref{tab:capacity}.
Note that all codec--bit-depth pairs satisfy $R \ge 1$ with sufficient margin for majority-vote decoding, achieving exact payload recovery under all manipulation conditions considered in this work.

\begin{table}[t!]
\caption{Number of repetitions ($R$) for each codec--bit-depth combination, computed for a $\seconds{1}$ carrier.}
\vspace{-0.12cm}
\centering
    \begin{tabular}{llccc}
        \toprule
        \multirow{2}{*}{Codec} & \multirow{3}{*}{\shortstack{Bitrate\\(kbps)}} & \multicolumn{3}{c}{Repetitions $R$ (per second)} \\
        \cmidrule(lr){3-5}
                               &       & LSB-1 & LSB-2 & LSB-4 \\
        \midrule
            SNAC               & 0.98  & 15    & 30    & 61    \\
            SemantiCodec       & 0.65  & 22    & 44    & 89    \\
            TAAE               & 0.40  & 18    & 36    & 74    \\
        \bottomrule
    \end{tabular}
    \vspace{\defvspace}
\label{tab:capacity}
\end{table}

\begin{table*}[t!]
\caption{Utterance-level detection EER (\%) $\downarrow$ and frame-level localization AUC$\uparrow$ for replacement and structural manipulations. Del., Ins.\ same, and Ins.\ diff.\ denote deletion, same-speaker insertion, and different-speaker insertion, respectively.}
\vspace{-0.06cm}
\centering
\resizebox{\textwidth}{!}{
\begin{tabular}{llcccccc ccccccc}
\toprule
& & \multicolumn{6}{c}{Detection EER (\%) $\downarrow$} & \multicolumn{6}{c}{Localization AUC $\uparrow$} \\
\cmidrule(lr){3-8} \cmidrule(lr){9-14}
& & \multicolumn{2}{c}{Replacement} & \multicolumn{3}{c}{Structural} &
  & \multicolumn{2}{c}{Replacement} & \multicolumn{3}{c}{Structural} & \\
\cmidrule(lr){3-4} \cmidrule(lr){5-7} \cmidrule(lr){9-10} \cmidrule(lr){11-13}
Codec & Method & Direct & TTS & Del. & Ins.\ same & Ins.\ diff. & Mean
               & Direct & TTS & Del. & Ins.\ same & Ins.\ diff. & Mean \\
\midrule
 \multirow{3}{*}{SNAC}
 & LSB-1   &  8.56 & 10.68 & 18.57 & 18.51 & 19.41 & 15.15 & 0.912 & 0.889 & 0.662 & 0.868 & 0.864 & 0.839 \\
 & LSB-2   &  8.80 & 10.61 & 18.17 & 18.39 & 19.05 & 15.00 & 0.912 & 0.889 & 0.662 & 0.868 & 0.864 & 0.839 \\
 & LSB-4   &  8.80 & 10.75 & 18.50 & 18.43 & 19.21 & 15.14 & 0.912 & 0.889 & 0.662 & 0.868 & 0.864 & 0.839 \\
\midrule
 \multirow{3}{*}{SemantiCodec}
 & LSB-1   &  8.40 &  9.78 & 17.47 & 18.13 & 18.55 & 14.47 & 0.911 & 0.892 & 0.661 & 0.868 & 0.862 & 0.839 \\
 & LSB-2   &  8.80 &  9.96 & 17.91 & 18.04 & 19.20 & 14.78 & 0.912 & 0.891 & 0.665 & 0.869 & 0.863 & 0.840 \\
 & LSB-4   &  8.08 &  9.97 & 18.05 & 18.12 & 19.30 & 14.70 & 0.912 & 0.893 & 0.662 & 0.867 & 0.864 & 0.840 \\
\midrule
 \multirow{3}{*}{TAAE}
 & LSB-1   & 28.43 & 29.11 & 36.20 & 33.50 & 33.32 & 32.11 & 0.871 & 0.841 & 0.614 & 0.802 & 0.799 & 0.785 \\
 & LSB-2   & 28.33 & 29.08 & 36.20 & 33.46 & 33.16 & 32.05 & 0.871 & 0.841 & 0.614 & 0.802 & 0.799 & 0.785 \\
 & LSB-4   & 28.36 & 29.04 & 36.16 & 33.46 & 33.20 & 32.05 & 0.871 & 0.841 & 0.614 & 0.802 & 0.799 & 0.785 \\
\bottomrule
\end{tabular}
}
\vspace{\defvspace}
\label{tab:results}
\end{table*}

\begin{figure}[t!]
    \centering
    \includegraphics[width=0.9\linewidth]{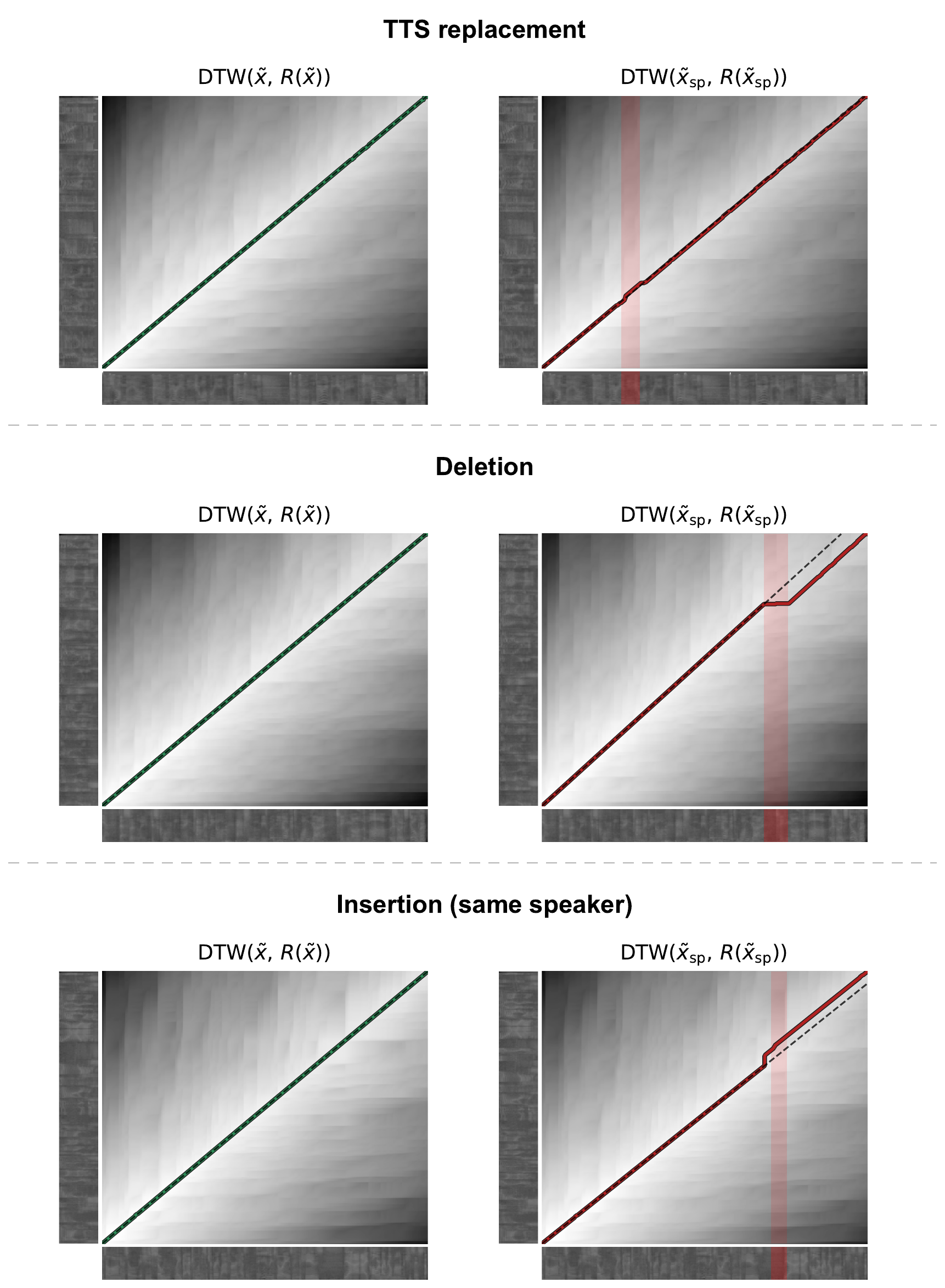}
    \vspace{-0.12cm}
    \caption{DTW cost matrices and optimal warping paths for each manipulation type. \textbf{Left:} alignment between the intact watermarked signal $\xwm$ and its self-reconstruction $R(\xwm)$, showing a near-diagonal path and low cost. \textbf{Right:} alignment between the manipulated watermarked signal $\xwmspoof$ and its self-reconstruction $R(\xwmspoof)$, with the manipulated region highlighted in red.}
    \label{fig:dtw_visualization}
    \vspace{\defvspace}
\end{figure}

\section{Manipulation Conditions and Dataset}
\label{sec:eval_conditions}
To assess the performance limits of the proposed framework under ideal conditions for proactive defense, we evaluate across four controlled manipulation types that cover 
qualitatively distinct ways in which a watermarked signal may be altered. 
No channel degradation, compression, or additive noise is applied; the evaluation is designed to characterize what accuracy is achievable in the absence of such factors.

\textbf{Direct replacement.}
One or more word-level segments of $\xwm$ are replaced with acoustically matched material drawn directly from authentic recordings of the same speaker, without any re-synthesis. 
This is a purely local manipulation detectable solely through payload mismatch.

\textbf{TTS replacement.}
Synthesized segments produced by zero-shot voice cloning systems\,\cite{KimKS21_VITS_ICML, CasanovaEtAl22_YourTTS_ICML} are substituted for the corresponding regions of $\xwm$, 
using GT word-level boundaries from AV-Deepfake1M annotations.

\textbf{Deletion.}
One word segment is removed from $\xwm$ without replacement. 
The resulting temporal shift displaces all subsequent audio, disrupting the alignment of every payload repetition following the deletion point.

\textbf{Insertion.}
A word segment from a donor recording, drawn from the same or a different speaker, is inserted at a gap between two consecutive words of $\xwm$. 
As for deletion, the induced temporal shift disrupts all subsequent payload repetitions.

All manipulations are applied at the waveform level via an overlap-add concatenation procedure ($\ms{6}$ frame shift, $\ms{12}$ frame length, two-frame smoothing buffer), 
following\,\cite{OezerEtAl26_AudioStegoPartialDeepfake_Interspeech}. 
Active speech levels are normalized; no codec or vocoder processing is applied.
As illustrated in Fig.~\ref{fig:dtw_visualization}, the four manipulation types produce qualitatively distinct DTW distortion patterns.
In the absence of manipulation the warping path closely follows the diagonal, replacement manipulations produce sharp local deviations, while deletion and insertion induce a global temporal shift in the warping path following the manipulation point.

\textbf{Dataset.}
Experiments are conducted on an evaluation set derived from the validation split of the AV-Deepfake1M benchmark\,\cite{CaiEtAl24_AVDeepfake1M_ACMMM}, retaining 
the $1{,}480$ utterances for which authentic recordings and Whisper ASR\,\cite{RadfordEtAl23_Whisper_ICML} word-level transcriptions are available. 
The validation split is used since GT labels and word-level metadata are publicly available only for training and validation partitions.
Donor recordings for insertion are drawn from the VoxCeleb2 development and test splits\,\cite{ChungNZ18_VoxCeleb2DeepSpeaker_Interspeech}. 
%

\section{Evaluation}
\label{sec:evaluation}
\subsection{Experimental Conditions}
\label{sec:exp_conditions}

\textbf{Neural codecs.}
SNAC, SemantiCodec, and TAAE are used at $\kbps{0.98}$, $\kbps{0.65}$, and $\kbps{0.4}$, respectively (see Table~\ref{tab:capacity}).
For SNAC, signals are resampled to $\khz{24}$ for codec encoding and decoding only, while watermarking embedding and all subsequent evaluation are performed at $\khz{16}$.

\textbf{LSB.} LSB embedding is evaluated for $k \in \{1, 2, 4\}$ bits per sample, with a $64$-bit synchronization preamble prepended to each payload repetition.

\textbf{DTW-based scoring.}
Detection and localization scores are derived from DTW alignment between the received signal and its self-reconstruction, computed on $40$-band log-mel spectrograms with a $\ms{16}$ frame shift.
Cosine distance is used as the local dissimilarity measure, and the per-frame cost profile is smoothed prior to localization evaluation.

\subsection{Evaluation Methodology}
\label{sec:baseline}
We evaluate utterance-level detection using the equal error rate (EER). 
The detection score is computed as the mean of the top $1\%$ of smoothed per-frame DTW cosine costs between the received signal and its self-reconstruction, computed as $\devdtw(\xwm, R(\xwm))$ for authentic signals and $\devdtw(\xwmspoof, R(\xwmspoof))$ for manipulated signals.
The decision threshold $\tau$ in \eqref{eq:detection} corresponds to the operating point at which the false positive rate equals the false negative rate.
Lower EER values indicate better discriminative capacity between authentic and manipulated signals.

For frame-level localization, performance is measured by the Area Under the Receiver Operating Characteristic (ROC) curve (AUC). 
The AUC is computed between the smoothed per-frame DTW cost profile and binary GT frame labels derived from word-level boundaries provided by the dataset annotations. A perfect localization performance yields an AUC of $1.0$, while an AUC of $0.5$ corresponds to random guessing.
Higher AUC values indicate better localization of manipulated regions.

\subsection{Results}
\label{sec:results}

\textbf{Imperceptibility.}
The watermarked signal $\xwm$ attains a wideband {PESQ}\,\cite{ITUT01_PESQ_P862} of $4.64$, the maximum attainable score, for all $k \in \{1, 2, 4\}$, confirming that the embedding remains perceptually transparent even at the highest bit depth.

\textbf{Hash-based upper bound.}
As a reference, we evaluate a baseline which embeds a cryptographic hash of each local audio segment using the same LSB carrier method.
Under ideal conditions, this baseline achieves near-perfect performance (detection EER near 0\%, localization AUC above 0.999 across all manipulation types).
This upper-bound performance is achieved because any localized modification results in a completely different decoded hash with high probability, making the detection of altered frames mathematically trivial.
Such schemes, however, authenticate content without describing it, and thus cannot recover the original speech.

\textbf{Detection and Localization.}
Across every codec, bit depth, and manipulation type, the embedded payload is recovered without bit errors after majority voting.
The reported results therefore reflect the discriminability of the mismatch score rather than payload recovery failures.
Table~\ref{tab:results} reports utterance-level EER and frame-level AUC for all codec--bit-depth combinations.
As anticipated, perfect accuracy is not achieved by any configuration, confirming that the codec-based approach trades the exactness of hash-based verification for recovery capability.
SemantiCodec achieves the lowest mean EER, ranging from $14.47\%$ to $14.78\%$, closely followed by SNAC at $15.00\%$ to $15.15\%$.
TAAE yields a considerably higher mean EER at $32.05\%$ to $32.11\%$.
Across all codecs, direct replacement is the most detectable manipulation, while deletion is consistently the hardest as it removes content without introducing external material.
Detection also improves with manipulation duration, as shorter manipulations produce mismatch scores overlapping with the authentic distribution.

The localization results follow the same trend: SNAC and SemantiCodec achieve closely comparable mean AUC of $0.839$--$0.840$ across all bit depths, while TAAE yields the lowest score of $0.785$.
Deletion is again the hardest manipulation to localize as removing content compresses the DTW path rather than producing a high-cost region.
Direct and TTS replacement yield the highest AUC, between $0.841$ and $0.912$.

\textbf{Recovery quality.}
\label{sec:recovery_quality}
Unlike classical hash-based schemes, which leave tampered regions completely unrecoverable, the proposed framework successfully reconstructs the authentic speech within manipulated regions at codec-level fidelity, with zero payload bit errors.
The receiver reconstructs $R(\tilde{x}_{sp}) = g(D(\tilde{x}_{sp}))$, successfully restoring the original spoken content.
The reconstructions attain a wideband PESQ of $1.75$, $1.63$, and $1.65$ for SNAC, SemantiCodec, and TAAE, respectively, on a scale from $1.0$ to $4.64$ where higher values indicate better perceptual quality.
While these moderate PESQ values reflect the sample-level waveform deviations inherent to ultra-low-bitrate neural compression, these values do not compromise the usability of the restored signal as the reconstructed speech remains fully intelligible, natural, and speaker-consistent.
Importantly, our analysis shows that reconstruction quality and detection performance are decoupled: SemantiCodec yields the lowest reconstruction PESQ but the lowest detection EER. 
Conversely, TAAE's transformer decoder prioritizes generating highly natural-sounding acoustic variations rather than maintaining sample-faithful alignment with the original source.
This generative behavior increases the baseline mismatch score on authentic signals, which ultimately degrades the statistical separation required for accurate detection and localization.

\section{Conclusion}
\label{sec:conclusion}
We presented training-free self-embedding audio watermarking framework with ultra-low-bitrate neural codecs, in which a compact codec representation of the carrier is embedded into the signal itself.
Across all LSB bit depth--codec configurations, the embedded payload is recovered without bit errors, so that the authentic content of manipulated regions can always be reconstructed at codec fidelity, with intelligible and natural-sounding recovered speech, which cannot be offered by hash-based verification.
Detection and localization, obtained from the mismatch between the received signal and its self-reconstruction, remain imperfect even under ideal conditions, with deletion posing the greatest challenge.
Our results indicate that performance is driven primarily by the codec's reconstruction characteristics, with LSB bit depth playing only a limited role.
Future work includes semi-fragile embedding for robustness under channel distortions and further payload compression to increase embedding redundancy.

\bibliographystyle{IEEEtran}
\bibliography{references_compressed}
\end{document}